\documentclass[showpacs,amsmath,nofootinbib,amssymb,reprint,prl]{revtex4-1}
\pdfoutput=1
\usepackage{mathtools}
\mathtoolsset{showonlyrefs}
\usepackage{psfrag}
\usepackage{array}
\usepackage{amssymb}
\usepackage{amsmath}
\usepackage{amsthm}
\usepackage{graphicx}
\usepackage{epstopdf}
\usepackage{color}
\usepackage{braket}
\usepackage{comment}

\def\be{\begin{equation}}
\def\ee{\end{equation}}
\def\bea{\begin{eqnarray}}

\def\eea{\end{eqnarray}}

\usepackage{epsfig}
\allowdisplaybreaks

\begin{document}
\preprint{}[CCTP-2017-8,ICTP-IPP 2017/19]

\title{Antipodal correlation on the meron wormhole and a Bang-Crunch universe}

\author{Panagiotis Betzios}
\email{pbetzios@physics.uoc.gr}
\affiliation{Crete Center for Theoretical Physics, Institute for Theoretical and Computational Physics, Department of Physics, University of Crete 71003
Heraklion, Greece.}

\author{Nava Gaddam}
\email{gaddam@uu.nl}
\affiliation{Institute for Theoretical Physics and Center for Extreme Matter and Emergent Phenomena, Utrecht University, 3508 TD Utrecht, The Netherlands.}

\author{Olga Papadoulaki}
\email{O.Papadoulaki@soton.ac.uk}
\affiliation{Mathematical Sciences and STAG Research Centre, University of Southampton, Highfield, Southampton SO17 1BJ, United Kingdom.}

\begin{abstract}
We present a covariant euclidean wormhole solution to Einstein Yang-Mills system and study scalar perturbations analytically. The fluctuation operator has a positive definite spectrum. We compute the Euclidean Green's function, which displays maximal antipodal correlation on the smallest three sphere at the center of the throat. Upon analytic continuation, it corresponds to the Feynman propagator on a compact Bang-Crunch universe. We present the connection matrix that relates past and future modes. We thoroughly discuss the physical implications of the antipodal map in both the Euclidean and Lorentzian geometries and give arguments on how to assign a physical probability to such solutions.
\end{abstract}

\pacs{}
\maketitle

\section{Introduction}
Euclidean wormholes~\cite{Giddings:1987cg,Lavrelashvili1987,Hawking1988} are extrema of the Euclidean action whose interpretation is still partially shrouded in mystery. Originally they were proposed as a resolution to the cosmological constant problem~\cite{Coleman1988a}, and as objects that lead to a loss of quantum coherence causing an inherent uncertainty in the fundamental constants of nature \cite{Lavrelashvili1987,Coleman1988,Giddings1988}. An alternative interpretation was given in \cite{Hawking1990} within the context of the Wheeler-de Witt equation. Besides several problems specific to these proposals, a basic general question that remains unanswered is whether such solutions are stable (and thus minima of the Euclidean action), or if they should alternatively be thought of as bounces (or maxima) that might nevertheless contribute to the path integral in some non-perturbative fashion akin to \cite{Basar2013}.

On a parallel note, defining quantum gravitational observables in closed universes is an acute problem \cite{Witten:2001kn} whose few proposed resolutions suffer from closed time-like curves and other intricacies~\cite{Elitzur:2002rt}. 

Inspired by these questions, we study a Euclidean meron wormhole \cite{Hosoya:1989zn} in light of the antipodal $\mathbb{Z}_2$ mapping proposed by 't Hooft. We find a positive definite spectrum for scalar perturbations, whose Euclidean Green's function exhibits large antipodal correlation localised near the smallest sphere at the center of the throat after performing the antipodal map. The analytic continuation of this solution results in a finite Bang-Crunch geometry \cite{Maldacena:2004rf}, with temporal and spatial boundaries, opening up a possible handle on the problem of observables.

\section{The Meron Wormhole \label{sec:wormholesol}}
Consider the Euclidean Einstein-Yang Mills system:
\begin{equation}\label{eucl}
S = \int d^4 x\sqrt{g} \left(-\frac{1}{16{\pi}G_N}R  +\frac{1}{4 }\left(F^{a}_{\mu\nu}\right)^2\right)
\end{equation}
in units $\hbar = c =1$, where $R$ is the Ricci scalar and $F^{a}_{\mu\nu}$, the field strength for the $SU(2)$ (possibly embedded in $SU(N)$) gauge field $A^{a}_{\mu}$. The field strength is defined as $F^{a}_{\mu\nu} \coloneqq {\partial}_{\mu} A^{a}_{\nu}-{\partial}_{\nu} A^{a}_{\mu} + g_{YM} {\epsilon}^{abc}A^{b}_{\mu}A^{c}_{\nu}$. In addition to the Yang-Mills equations of motion $D^{\mu} F^{a}_{\mu\nu} = 0$ and the corresponding Bianchi identities, the Einstein equations of motion are given by
\begin{align}
R_{\mu\nu}-\frac{1}{2}g_{\mu\nu}R ~ &= ~ 8\pi G_N T_{\mu\nu}
\end{align}
where $R_{\mu\nu} = R^{\lambda}_{\mu\lambda\nu}$ and $T_{\mu\nu} = F^{a}_{\mu\rho} {F^{a}_\nu}^\rho - \frac{1}{4} g_{\mu\nu} \left(F^{a}_{\mu\nu}\right)^2$. The meron configuration $A_\mu^a = \eta_{a \mu \nu} x_\nu g^{-1}_{YM} \textbf{x}^{-2}$ yields the field strength
\begin{align}
F^a_{\mu\nu} = ~ \dfrac{1}{g_{\text{YM}}} \left[\eta_{a \mu \nu} \dfrac{f_1}{\textbf{x}^2} + \left(x_\mu \eta_{a \rho \gamma} x_\gamma - x_\rho \eta_{a \mu \gamma} x_\gamma\right)\dfrac{f_2}{\textbf{x}^2}\right] \, ,
\end{align}
with $f_1=f_2=-1$. Here, $\eta_{a \mu \nu}$ are the 't Hooft Eta symbols \cite{tHooft:1976snw}\textemdash whose conventions we follow\textemdash that mix the generators of space-time $SO(4)$ with those of the $SU(2)$ gauge symmetry. The summation over all lower indices is carried out by the flat metric $\delta_{\mu\nu}$; however, owing to the conformal nature of the Yang-Mills equations, this is a solution in any conformally flat background. The meron configuration was one of the first proposed mechanisms for confinement~\cite{Callan:1977qs,Lenz:2003jp}.
This meron-pair solution has a unit topological instanton charge split equally between the two points at $\textbf{x}=0$ and $\textbf{x}=\infty$; the pair may be brought closer together with a conformal transformation \cite{Callan:1978bm}. Unlike the BPST instanton \cite{Belavin:1975fg,tHooft:1976snw,Callan:1976je}, this solution is not self-dual. The meron's magnetic field lines provide the required locally-negative energy density to support a wormhole throat and source the right hand side of the Einstein equations which are solved by the metric
\begin{align}\label{eqn:euclideanmetric}
\mathrm{d}s^2 ~ &= ~ \left(1+ \frac{r_1^2}{\textbf{x}} \right)^2 \delta_{\mu \nu} \mathrm{d}x^\mu \mathrm{d}x^\nu \, \quad r^2_1 = \pi G_N g^{-2}_{YM}\, , \nonumber \\
\mathrm{d}s^2 ~ &= ~ \mathrm{d}r^2 + \left(r^2 + 4 r^2_1\right) \mathrm{d}\Omega^2_3
\end{align}
with $\textbf{x} = \sqrt{x_\mu x^\mu} \geq 0$ and $r = \textbf{x} - r^2_1 \textbf{x}^{-1}$. This transformation maps the regions $\textbf{x} \lessgtr r_1$ to the two sides of the wormhole $r \lessgtr 0$ and the smallest sphere to $\textbf{x}=r_1$. The geometry is that of a two-sided Euclidean space (with the two asymptotic regions at $r = \pm \infty$) connected by a wormhole throat of smallest size $2 r_1$; it has the curious property that the regime of its semi-classical validity coincides with the perturbative regime of the YM theory.  The metric admits a natural analytic continuation $r = i t$ into a Big Bang - Big Crunch universe, with singularities at $r= \pm 2 r_1$~\cite{Maldacena:2004rf}. It is an open problem to find general multi-meron solutions specified by functions $f_1\left(\textbf{x}^2\right)$ and $f_2\left(\textbf{x}^2\right)$, constrained by both Einstein and Yang-Mills equations.

Let us mention here that the Euclidean action is logarithmically divergent. This is easy to see since the Ricci scalar vanishes ($R=0$) and, from the Yang-Mills Lagrangian, one finds the usual log-divergence of the meron configuration $\sim \log(L/a)$ where $L$ is the system size (IR-cutoff) and $a$ is a UV-cutoff~\cite{Callan:1977gz}. This problem has been discussed in the past, for example in~\cite{Hosoya:1989zn} and later in~\cite{Gupta:1989bs}. The UV problem is evaded by the finite size of the throat. The IR divergence can be evaded by considering similar solutions in the presence of a small cosmological constant that provides a natural regulator for such IR divergences. This is one possible and known way of making sense of similar configurations; however, a satisfactory solution for asymptotically flat space-times is not known. Inspired by the work of Kosterlitz and Thouless~\cite{Kosterlitz1973} (whose relevance to merons is sketched in \cite{Callan:1977gz}), we propose that the seemingly infinite classical action must be compared with the entropy of such configurations. A known issue with the pure meron background~\cite{Callan:1977gz} is that one needs to regulate the meron core with an ad-hoc procedure which results in a violation of the equations of motion for the regulated configurations. This UV problem is evaded by the meron wormhole owing to its finite-sized throat, a fact that will become even more transparent in our novel interpretation\textemdash using an antipodal identification\textemdash of the wormholes as holes of nothing.

\paragraph{Symmetries of the background} Flat space has ten killing vectors (KVs) and five conformal KVs (CKVs). Since the wormhole is conformally flat, its symmetries may be found by studying if the former remain KVs and if any of the latter get promoted to KVs. 
It may be checked that while rotations remain killing vectors of the wormhole background, translations become CKVs. Additionally, none of the five CKVs (dilatation and special conformal transformations) of flat space are promoted to KVs. Finally, the wormhole background possesses a discrete inversion symmetry under $x^\mu \rightarrow r^2_1 x^\mu \textbf{x}^{-2}$. In the $r$ coordinates, it reads $r\rightarrow -r$. This is an antipodal mapping when appended with appropriate maps on the three-sphere, similar to the one considered in \cite{Hooft:2016itl,Betzios:2016yaq} in the context of 't Hooft's black hole S-Matrix. 

\section{Scalar perturbations\label{sec:scalarpert}}
The action governing the fluctuations of a charged scalar field in this background, is given by $S^{(2)}_\phi = \int \mathrm{d}^4 x \, \sqrt{g} \, \mathcal{L}^{(2)}_\phi \equiv  \int  \mathrm{d}^4 x \, \sqrt{g} \phi^\star \mathcal{M}_\phi \phi $ with
\begin{align}\label{eqn:scalaraction}
\mathcal{L}^{(2)}_\phi =   \left[\left|D_\mu \phi\right|^2 + \left(m^2 + \zeta R\right) \left|\phi\right|^2\right] \, ,
\end{align}
where $D_\mu \phi \eqqcolon \left(\nabla_\mu + i g_{\text{YM}} A_{\mu,a}\right)\phi$. In order to explicitly evaluate the spectrum of the operator $\mathcal{M}_\phi$, it is first useful to define the following space-time operators, which represent rotations in the two invariant SU(2) subgroups of the rotation group SO(4): 
\begin{equation}
L^a_1 \coloneqq - \dfrac{1}{2} i \eta_{a \mu \nu} x_\mu \partial_\nu \quad \text{and} \quad L^a_2 \coloneqq - \dfrac{1}{2} i \bar{\eta}_{a \mu \nu} x_\mu \partial_\nu \label{eqn:Loper12}
\end{equation}
with $\left[L^a_p , L^b_q\right] = i \delta_{pq} \epsilon^{abc} L^c_p$ and ${L_1}^2 = {L_2}^2 = L^2  = - \left(1/8\right) \left(x_\mu \partial_{\nu} - x_\nu \partial_{\mu}\right)^2$. Owing to the $SU(2)$ projection of these operators, the eigenvalues of $L^2$ are given in terms of half-integers $l$ via $L^2 = l \left(l+1\right)$. Furthermore, for scalars, isospin rotations are generated by the operators $T^a$ with eigenvalues $T^2 = t \left(t+1\right)$, with an arbitrary total isospin $t$. The operator $\mathcal{M}_\phi$  (whose vacuum counterpart we call $\mathcal{M}_0$) can now be written as 
\be\label{eqn:operator}
\mathcal{M}_\phi = -\nabla^2 + \dfrac{\textbf{x}^2 \left[T^2 + 4 \left(T\cdot L^a_1\right) \right]}{\left(\textbf{x}^2 + r^2_1\right)^2}  + m^2  \, , 
\ee 
where $\nabla^2$ is the Laplace-Beltrami operator of the wormhole background.
In order to calculate the spectrum (of the vacuum and the wormhole background), it turns out to be convenient to solve the rescaled operator equation $\mathcal{V}_{\phi,0} \phi = \lambda r_1^{-2} \phi$, with $\mathcal{V}_{\phi,0} \coloneqq \mathcal{A} \mathcal{M}_{\phi,0} \mathcal{A}\, , \, \mathcal{A}= \left(\textbf{x}^2 + r^2_1\right)^2 /\left(4 r_1^2 \textbf{x}^{2}\right) $ \cite{tHooft:1976snw}. This corresponds to solving the Schr\"{o}dinger equation
\begin{widetext}
\begin{align}\label{eqn:schrodinger1}
\left[\partial^2_\textbf{x} + \left(\dfrac{3}{2 \textbf{x}} - \dfrac{2 r^2_1}{\textbf{x} \left(\textbf{x}^2 + r^2_1\right)}\right) \partial_\textbf{x} - \dfrac{J^2_1}{\textbf{x}^2} + \dfrac{16 r^2_1 \left(\lambda - r_1^2 m^2\right)}{\left(\textbf{x}^2 + r^2_1\right)^2}\right] \phi ~ &= ~ 0 \, ,
\end{align}
\end{widetext}
where $J_1 \coloneqq T + 2 L_1$. In order to solve this equation, we expand the field in three-sphere harmonics that satisfy $-\nabla^2_{S^3} Y_{k\tilde{l}\tilde m}(\Omega) = k\left(k+2\right) Y_{k\tilde{l} \tilde m}(\Omega)$ with $\left(k+1\right)^2$-fold degeneracy since $\tilde m = -\tilde{l}, \dots, \tilde{l}$ and $\tilde{l} = 0, 1, \dots, k$. Similarly, $J^2_1 Y_{k\tilde{l} \tilde m}(\Omega) = j_1\left(j_1+2\right) Y_{k\tilde{l} \tilde m}(\Omega)$ with $j_1 = k-t, \dots, k+t$ provided $j_1 \geq 0$. The Schr\"{o}dinger equation \eqref{eqn:schrodinger1} then reduces to a one-dimensional problem whose solutions are associated Legendre polynomials. To bring them into canonical form, we redefine the field $\phi$ as $\phi = \left(1 - y^2\right)^{1/2} \psi/\left(2r_1\right)$ and change variables as $y = \left(r^2_1-\textbf{x}^2\right)/\left(r^2_1 + \textbf{x}^2\right)$. In these coordinates, the two boundaries $\textbf{x}=0,\infty$ are at $y=\pm 1$ while the smallest sphere $\textbf{x}=r_1$ lies at $y=0$. As it does on the $r$ coordinate, the discrete $\mathbb{Z}_2$ map acts as a reflection $y \rightarrow -y$ on this coordinate. The Schr\"{o}dinger equation \eqref{eqn:schrodinger1} then reduces to the canonical associated Legendre form 
\begin{align}\label{eqn:schrodinger2}
\left[\left(1 - y^2\right) \partial^2_z - 2 y \partial_z + \left(\nu \left(\nu +1\right) - \dfrac{\mu^2}{1 - y^2}\right)\right] \psi = 0 \quad
\end{align}
whose two linearly independent solutions are $P^\mu_\nu\left(y\right)$ and $Q^\mu_\nu\left(y\right)$ with $\nu = \frac{1}{2} \left(\sqrt{1 + 16 \lambda - 16 r_1^2 m^2} -1\right)$ and $\mu = j_1 + 1$. Since the order $\mu$ is an integer, the normalisable solutions (in the domain $y\in \left[-1,1\right]$) to this equation are given by demanding that the degree $\nu$ also be an integer. Therefore, for the spectrum, we find 
\begin{align}
\lambda_{n,j_1} - r_1^2 m^2 ~ = ~ \dfrac{1}{4}\left(n^2 + n\right) \, , \quad n \geq j_1 + 1 \, .
\end{align}
Of immediate note is the fact that the spectrum is positive definite and the wormhole stable under (Gaussian) scalar perturbations. Furthermore, since the meron wormhole breaks all the non-compact symmetries of flat space, there are no zero modes either. The vacuum spectrum is calculated most easily in the original $r$ coordinates~and the vacuum regulated one-loop determinant of scalar fluctuations $\det \mathcal{M}_\phi / \det \mathcal{M}_0 $ can be calculated following the prescription in~\cite{tHooft:1976snw}. Instead, in what follows we turn to a calculation of the Green's function for the scalar perturbations.

\section{Correlation function\label{sec:correlator}}

The Euclidean Green's function can be written in terms of the homogeneous solutions ($\lambda = 0$) to \eqref{eqn:schrodinger2} and the Wronskian determinant $W\left(P^\mu_\nu,Q^\mu_\nu\right) = P^\mu_\nu\left(x\right) Q'^\mu_\nu\left(x\right) - P'^\mu_\nu\left(x\right) Q^\mu_\nu\left(x\right)$, with a prime indicating a derivative with respect to the argument $x$. This two-point function with homogeneous boundary conditions for the field $\psi$ is given by
\begin{widetext}
\begin{equation}\label{eqn:euclideangreen}
G^E\left(x,\Omega;y,\Omega'\right) ~ = ~ \sum_{k=0}^{\infty}\sum_{\mu = k-t+1}^{k+t+1} \sum_{\tilde{l}=0}^{\mu-1}\sum_{\tilde m=-\tilde{l}}^{\tilde{l}}  \left( \dfrac{\Theta\left(y-x\right) P^\mu_\nu\left(x\right) Q^\mu_\nu\left(y\right) + x \leftrightarrow y }{\left(1 - y^2\right) W\left(P^\mu_\nu,Q^\mu_\nu\right)\left(y\right)}  \right) Y_{\left(\mu-1\right) \tilde{l} \tilde m}\left(\Omega\right)Y_{\left(\mu-1\right) \tilde{l} \tilde m}\left(\Omega'\right) \, .
\end{equation}
\end{widetext}
The corresponding correlator for $\phi$ is achieved by the rescaling that relates the fields. First we perform the sum over the indices $\tilde{l}$ and $\tilde m$ and then the sum over $\mu$ using addition theorems~\cite{JeffreyZwillinger_GradRyzh}. The result for $t=0$ is
\begin{align}\label{eqn:massiveGE}
G^E\left(x,\Omega;y,\Omega'\right) = - \dfrac{\sqrt{\left(1- x^2\right) \left(1-y^2\right)}}{4 \pi^2 \sqrt{1 - \xi^2}} Q^1_\nu \left(\xi\right)
\end{align}
where $\xi = x y + \sqrt{\left(1- x^2\right) \left(1-y^2\right)} \cos\left(\delta \Omega\right)$ and $\cos\left(\delta \Omega\right)$ is the angular difference, on the three-sphere, between $\Omega$ and $\Omega'$. A non-vanishing isospin requires an extra finite summation that can be performed but is not presented here. Taking the massless limit $m^2 \rightarrow 0$, we find
\begin{equation}\label{eqn:masslessGE}
G^E_{m^2=0} \left(x,\Omega ; y, \Omega'\right) ~ = ~ \dfrac{\sqrt{\left(1- x^2\right) \left(1-y^2\right)}}{4 \pi^2 \left(1+\xi\right) \left(1-\xi\right)} \, .
\end{equation}
As it must, the function displays maximal correlation when $\left(x,\Omega\right)$ coincides with $\left(y,\Omega'\right)$. Rather beautifully, however, we find maximal correlation in equal measure when $\left(y,\Omega'\right) = \left(-x,\bar{\Omega}\right)$ is the reflected-antipodal point of $\left(x,\Omega\right)$, allowing for a natural $\mathbb{Z}_2$ identification between the two sides of the wormhole $\left(y, \Omega\right) \equiv \left(-y,\bar{\Omega}\right)$. This is further corroborated by the fact that modes on either side are related by a Unitary mapping through Legendre function connection formulae \cite{JeffreyZwillinger_GradRyzh}. After the identification, the correlation between antipodal points, on the same side of the wormhole, is a local effect near the smallest sphere. This may be seen from the following observation: With $x, ~ y \geq 0$, and $\Omega' = \bar{\Omega}$, the antipodal point of $\Omega$, the correlation is maximal when $x=y=0$. That is when they lie on the smallest sphere. As we move away from the center of the throat into the single exterior, the correlation on antipodal points of the three sphere diminishes 
, rendering the maximal entanglement a local effect near the center of the throat.  This is a concrete realisation of the antipodal identification suggested in \cite{Hooft:2016itl,Betzios:2016yaq}. As in that case, the spectrum is naturally halved owing to the following relation between three-sphere harmonics on antipodal points: $Y_{j_1 \tilde{l} \tilde m}\left(\bar{\Omega}\right) = \left(-1\right)^{j_1}Y_{j_1 \tilde{l} \tilde m}\left(\Omega\right)$. 

The $x, y$ dependence of the massless correlator \eqref{eqn:masslessGE} can be entirely written in terms of a cross-ratio $\left[\left(y_3 - y_1\right)\left(y_4 - y_2\right)/\left(y_3 - y_2\right)\left(y_4 - y_1\right)\right]^{1/2}$ with $y_{1,2,3,4} = \left(y,x,1/x,1/y\right)$. This cross-ratio is invariant under the entire projective linear group of M\"{o}bius transformations. Moreover, the $\mathbb{Z}_2$ mapping renders the manifold non-orientable. It is curious to note that this is reminiscent of string theory which admits such vacua, in the presence of orientifolds, that contribute to the path integral. 

\begin{figure}[t]
\includegraphics[width=80mm]{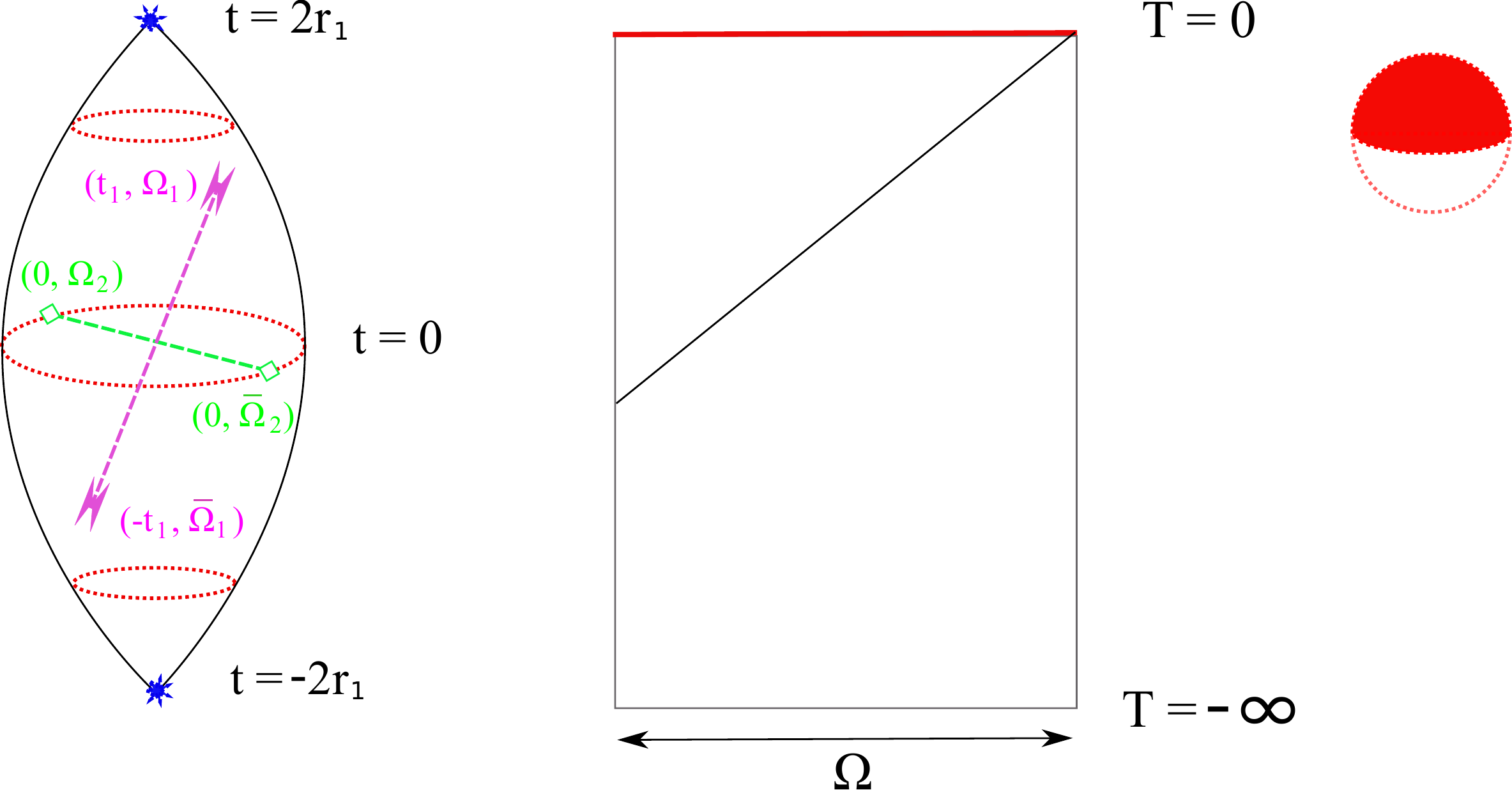}
\caption{Left: The Bang-Crunch universe in real time with $\Omega_i$ being antipodal to $\bar{\Omega}_i$. Right: Antipodal identification results in half of the Einstein static universe in conformal time, with half a three sphere (whose coordinates are collectively represented by $\Omega$) at the future boundary $T=0$. The solid line inside is null.}
\label{fig:geometry3}
\end{figure}

Having discussed the antipodal mapping, we now return to the issue of the divergence of the Euclidean action.\footnote{A similar argument might also work for the usual interpretation of meron wormholes but would be more involved.} As we saw above, after the said antipodal map, one is left with flat space from which a ball of radius $r_1$ is excised. Let us now consider an arbitrary Euclidean spacetime region of volume $L^4$. This hole with a rough ``core volume'' of $r_{1}^{4}$ could be centered at any spacetime point in this volume. The probability of such a configuration appearing is thus $(L/r_1)^4 e^{-c \log L/r_1 },$ (with $ c \sim 1/ g_{YM}^{2}$) containing an ``entropic" pre-factor and of course, the action  \cite{Rey:1989th}. Therefore, when computing the free energy, this entropic term competes against the on-shell action. For appropriate $g_{YM}$, the entropic term may dominate the classical action signalling a phase where the previous wormholes (now holes of nothing) proliferate. A transition between such a phase and the usual phase of gravity would be analogous to the unbinding transition of vortices in the Kosterlitz-Thouless example~\cite{Kosterlitz1973}. Furthermore, since $r_{1}^{2} \sim G_{N}/g_{YM}^{2}$, the result also depends on the running of the gravitational coupling constant. Of course this argument is incomplete, and one needs to perform a complete one loop computation of fluctuations on such a background because it directly affects the entropic pre-factor. Since this would be an extremely interesting phase of gravity, we intend to address such a possibility in full detail in future work~\cite{BGP}.

\section{Big-Bang Big-Crunch universe}
The euclidean meron-wormhole metric \eqref{eqn:euclideanmetric} yields a closed Big-Bang Big-Crunch universe upon analytically continuing $r \rightarrow i t = 2 i r_1 \tanh T$:
\begin{align}
\mathrm{d}s^2 ~ &= ~ - \mathrm{d}t^2 + \left(4 r^2_1 - t^2\right)\mathrm{d}\Omega^2_3 \, , \quad t \in \left[-2r_1, 2r_1\right] \\
&= ~ \dfrac{4 r^2_1}{\cosh^2 T}\left(-\mathrm{d}T^2 + \mathrm{d}\Omega^2_3 \right) \, , \quad T \in \left(-\infty,\infty\right) \, .
\end{align}
The end points of this closed universe represent Bang and Crunch singularities. This geometry is depicted in FIG. \ref{fig:geometry3}. The early universe begins to expand at $t=-2r_1$ and after reaching maximum size at $t=0$, collapses into a Crunch at $t=2r_1$. In conformal time, the geometry captures the \textit{complete} Einstein static universe.

Defining $\tilde{\mu}=\nu+\frac{1}{2}$ and $\tilde{\nu}=\mu-\frac{1}{2}$, a good basis of solutions is $P^{\pm \tilde{\mu}}_{\tilde{\nu}} (\tanh T)$. When $\tilde{\mu}$ is real, the solutions blow up at $T\rightarrow\pm\infty$ and result in an instability. For purely imaginary $\tilde{\mu}$ (that is, for $m^2>1/16$), the solutions oscillate in conformal time as $\exp\left(\pm i |\tilde{\mu}| T\right)$ or $\exp\left(\mp i |\tilde{\mu}| T\right)$ as $T\rightarrow \pm \infty$.  Moreover, the modes are related by complex conjugation $P^{+ \tilde{\mu}}_{\tilde{\nu}}(\tanh T) = [P^{- \tilde{\mu}}_{\tilde{\nu}}(\tanh T)]^*$. Therefore, the scalar field can consistently be quantised via $\psi(T, \Omega) =  \sum_{j_1, \tilde{l}, \tilde{m}} a_{j_1 \tilde{l} \tilde{m}} P^{\tilde \mu}_{j_1 + 1/2} (\tanh T) Y_{j_1 \tilde{l} \tilde{m}}(\Omega) + a_{j_1 \tilde{l} \tilde{m}}^\dagger P^{-\tilde{\mu}}_{j_1 + 1/2} (\tanh T) Y_{j_1 \tilde{l} \tilde{m}}(\Omega)$, with the creation and annihilation operators obeying canonical commutation relations. The vacuum is then defined by $a_{j_1 \tilde{l} \tilde{m}}\ket{0} = 0$. The expanding modes $P^{\pm \tilde{\mu}}_{\tilde{\nu}} \left(-\tanh T\right)$ may be obtained from the contracting ones $P^{\mp \tilde{\mu}}_{\tilde{\nu}} \left(\tanh T\right)$ via the connection matrix
\begin{equation}
\begin{pmatrix}
\alpha & \beta \\
\beta^* & \alpha^*
\end{pmatrix} = \begin{pmatrix}
\frac{\Gamma\left(\tilde{\nu} + \tilde{\mu} + 1\right)}{\Gamma\left(\tilde{\nu} - \tilde{\mu} +1\right)} \frac{\sin \pi \left(\tilde{\mu} + \tilde{\nu}\right)}{\sin \left(\pi \tilde{\mu}\right)} & - \frac{\sin \left(\pi\tilde{\nu}\right)}{\sin \left(\pi\tilde{\mu}\right)} \\
- \left[\frac{\sin \left(\pi\tilde{\nu}\right)}{\sin \left(\pi\tilde{\mu}\right)}\right]^* & \left[\frac{\Gamma\left(\tilde{\nu} + \tilde{\mu} + 1\right)}{\Gamma\left(\tilde{\nu} - \tilde{\mu} +1\right)} \frac{\sin \pi \left(\tilde{\mu}+ \tilde{\nu}\right)}{\sin \left(\pi \tilde{\mu}\right)} \right]^*
\end{pmatrix}~
\end{equation}
This is a canonical matrix with unit determinant, thereby gaining an interpretation of a Bogolyubov matrix in $T$ coordinates. Relative probability of particle creation in a given mode per unit volume can now be calculated to be $\left|\beta\right|^2\left|\alpha\right|^{-2} = \mathrm{sech}\left(\pi \tilde{\mu}\right)$. This is similar to the corresponding result in de-Sitter space \cite{Mottola:1984ar}. 

\section{Quantum observables and holography}
Big-Bang Big-Crunch universes such as this pose a serious problem of defining consistent quantum observables in the theory. While AdS has a conformal boundary at spatial infinity where correlators may be defined, the future and past null infinity of flat space allow for an S-matrix to be defined. While dS possesses a future boundary in conformal time, the case at hand is more difficult for there is neither a boundary in space, nor in time. Nevertheless, there is a possible resolution to this issue arising from the antipodal correlation between modes at either side of zero time.

Since the antipodal mapping acts on the coordinates $x,y$ just as it acts on the coordinate $r$\textemdash as a reflection\textemdash the Green's function \eqref{eqn:masslessGE} may easily be analytically continued by $x\rightarrow i x$ and $y \rightarrow iy$. It may be checked that the maximal antipodal correlation on either side of $T=0$ survives this continuation, rendering the $\mathbb{Z}_2$ mapping legitimate. Upon this identification, one obtains a Big-Bang universe naturally possessing a future boundary in time, at $t=T=0$; this is a spatial section of the geometry that is an antipodally identified three-sphere. Should a holographic dual exist, we would expect it to have a finite Hilbert space of states, owing to the compactness of the geometry. This boundary could then provide for a natural habitat for definition of observables in the quantum theory in a similar fashion to the dS proposed dualities. 

Finally, it is worth mentioning that at $t=0$, half of a euclidean meron wormhole may be glued to extend space-time. This may be done in two distinct ways. One is by gluing the smallest three-sphere  of the euclidean wormhole to the largest one of the expanding universe. Another is by gluing together antipodally identified three-spheres of the euclidean and time-dependent geometries at $r=0$ and $t=0$ respectively. Of course, additional possibilities arise when AdS and dS versions of the euclidean meron wormholes of \cite{Hosoya:1989zn,Maldacena:2004rf} are considered.
 
\section{Conclusions}

In this article we have reinterpreted euclidean wormholes as ``holes of nothing"; after the antipodal map, the single exterior is that of flat space from which a sphere of size $2r_1$ has been excised. We analytically computed the spectrum and the two-point function of scalar fluctuations on a Euclidean meron wormhole. The solution is found to be stable under probe scalar perturbations. We showed that the Green's function exhibits large antipodal correlation near the center of the throat. This gives good reason to thoroughly study the stability \cite{BGP} of this solution under generic metric and gauge field perturbations. Should there be negative modes as predicted in \cite{Giddings:1987cg,Gupta:1989bs,Maldacena:2004rf,Rubakov:1996cn}, it would be interesting to see if the antipodal identification projects them out. Returning to the logarithmic divergence of the classical Euclidean action, we note that the solution we use is naturally UV regulated by the size of the hole of nothing (after antipodal identification). We can nevertheless make sense of such solutions by comparing the euclidean action to their entropy in a given volume of space.
For a complete semiclassical treatment though, one needs to study all the field fluctuations on the background up to one-loop. Technically one computes a complete one loop determinant with the modes split in terms of their spin and where ghost degrees of freedom are properly introduced to account for gauge invariance. This calculation has not been performed for a reason, in the literature. The Yang-Mills fluctuations are coupled to the gravitational ones on such a background. This mixing results in an intricate calculation. In on-going work~\cite{BGP}, we find that this task is technically involved but possible to achieve. For instance, we find that spin-2 fluctuations have no negative modes and are stable. We hope to report on the complete 1-loop computation in the near future \cite{BGP}.

On another note, in AdS, finding large correlation between the two exteriors (as we found in this work) could address several CFT puzzles raised in \cite{Maldacena:2004rf}. It would also be interesting to revisit instabilities of other euclidean wormholes in this light. The antipodal map's role in confinement is worth exploring in the light of centre symmetry in SU(N) YM theory. It is also of importance to further investigate the `wrong-sign' problem of the conformal mode of the metric upon the background coupling to the longitudinal mode of the gauge field, perhaps also in conjunction with the antipodal identification; this may provide interesting constraints on the swampland in view of the weak gravity conjecture.

\section*{Acknowledgments}
We take pleasure in thanking Bernard de Wit, Umut G\"ursoy, Miguel Montero, Kostas Skenderis, Phil Szepietowski and especially Elias Kiritsis and Gerard 't Hooft for valuable discussions and suggestions. We also wish to thank the anonymous referee for comments and suggestions. PB is supported by the Advanced ERC grant SM-grav, No 669288. OP is supported by the STFC Ernest Rutherford grants ST/K005391/1 and ST/M004147/1.

\bibliography{references}
\end{document}